\documentclass[12pt]{amsart}
\usepackage{amsmath}
\usepackage{amscd}
\usepackage{amsfonts}
\usepackage{amssymb}
\usepackage{amsthm}

\newcommand{\T}{\mathbb{T}}
\newcommand{\R}{\mathbb{R}}
\newcommand{\Z}{\mathbb{Z}}
\newcommand{\Q}{\mathbb{Q}}

\newcommand{\Op}{\operatorname{Op}}
\newcommand{\OpN}{\operatorname{Op}_N}

\newcommand{\Cinf}{C^{\infty}}

\newtheorem{thm}{Theorem}[section]
\newtheorem{lem}[thm]{Lemma}%[subsection]
\newtheorem{cor}[thm]{Corollary}%[subsection]

\newtheorem{defn}{Definition}[section]

\newtheorem{remark}{Remark}[section] % \renewcommand{\theremark}{}

\begin{document}

\title[Quantum Unique Ergodicity for maps on the torus]{Quantum Unique Ergodicity for maps on the torus}%
\author{Lior Rosenzweig}%
\address{Tel Aviv University}%
\email{rosenzwe@post.tau.ac.il}%

\begin{abstract}
  When a map is classically uniquely ergodic, it is expected that its quantization
  will posses quantum unique ergodicity.
   In this paper we give examples of Quantum Unique
 Ergodicity for the perturbed Kronecker map,
  and an upper bound for the rate of
 convergence.
 \end{abstract}

\maketitle

\section{Introduction}
 \subsection{Background}
 One of the problems in Quantum Chaos is the asymptotic behavior
 of the expectation value in eigenstates. When quantizing classical dynamics on a phase space one constructs a Hilbert space of states ,$\mathcal{H}_h$,
 and an algebra of operators , the algebra of "quantum observables", that assigns for each smooth function on the phase space $f$ an operator
 $\Op_h(f)$ where $h$ implies dependence on Planck's constant $h$, and the dynamics is quantized to a unitary time evolution operator, $U_h$ on $\mathcal{H}_h$.
 For any orthonormal basis of eigenfunctions of $U_h$, $\{\psi_j\}$, the expectation value of $\Op_h(f)$ in the eigenstate $\psi_j$ is given
 by $\langle\Op_h(f)\psi_j,\psi_j\rangle$. The semiclassical limit of these is the limit where $h\to 0$.
 When the classical dynamics of a system is ergodic, it is known that the time average of the trajectories of the system converges
 to the space average. An analogue of this is given by Schnirelman's  Theorem \cite{Shn},\cite{Z1},\cite{C}, which states that for an ergodic system the
 expectation values of $\Op(f)$ converges to the phase space average of $f$, for all but possibly a zero density subsequence of
 eigenfunctions. This is referred to as quantum ergodicity. The case where there are no exceptional subsequences is referred to
 as quantum unique ergodicity (QUE).\\
\par
 When the phase space is $\T^2=\R^2/\Z^2$ it is required that each state will be periodic in both position and momentum and thus
 Planck's constant is restricted to be an inverse of an integer $h=\frac{1}{N}$, and the Hilbert space is of dimension $N$, namely
 $L^2(\Z/N\Z)$. The semiclassical limit in this case is the limit where $N\to\infty$. Given a continuous map $A$ on $\T^2$, we define its quantization
 as a sequence of unitary operators on $L^2(\Z/N\Z)$, $U_N(A)$ satisfying
  \begin{equation}\label{eq:egorov}
   \|U_N(A)^{-1}\OpN(f)U_N(A)-\OpN(f\circ A)\|\longrightarrow 0 \qquad \mbox{as } N\to \infty
  \end{equation}
 for all $f\in\Cinf(\T^2)$, where $f\circ A(p,q)=f(A(p,q))$. This is an analogue of Egorov's Theorem, and the eigenfunctions of
 $U_N(A)$ are analogues of eigenmodes.
  \par
 A first example of QUE was given on the 2-torus $\T^2$,by Marklof and Rudnick \cite{MR}, where the classical dynamics
 is an irrational skew translation, that is  classically uniquely ergodic.
 For this map they found that for generic translations, the rate
 of convergence is $O(N^{\frac{1}{4}+\epsilon})$. A famous example
 of a quantization of a map
 is of linear automorphism of $\T^2$ called the "CAT map",(\cite{HB},\cite{DEGI}), that is
 if $A\in SL(2,\Z)$. If $|tr A|>2$ that is if $A$ is hyperbolic,
 then the map is known to be ergodic, but not uniquely ergodic. In
 this case it was shown that there is no QUE (\cite{DebFaNon}),
 but there exists a special basis (Hecke Basis) for which QUE
 holds (\cite{KR}). In this case the rate of convergence was shown
 to be $O(N^{\frac{1}{4}+\epsilon})$, and is conjectured to be
 $O(N^{\frac{1}{2}+\epsilon})$. (It was shown that in the case where $N=p$ where $p$ is a prime number the rate of convergence is $O(p^{1/2})$ \cite{GH}).

 In this paper we will give a family of more examples of QUE on the 2-torus, all of them are also classically uniquely ergodic, and study the rate of convergence.\\

 \subsection{QUE for maps on the torus}
 The map in this paper will be the perturbed Kronecker map, that
 is
   \begin{eqnarray*}
   \Phi^\alpha_V:& \T^2 \to    &\T^2\\
    \Phi^\alpha_V :&
                    \begin{pmatrix}
                     p \\
                     q
                    \end{pmatrix}
                   \mapsto &
                    \begin{pmatrix}
                     p+\alpha_1 \\
                     q+\alpha_2+V(p)
                    \end{pmatrix} \mod 1
  \end{eqnarray*}
where $\alpha=(\alpha_1,\alpha_2)$, and $V(p)$ is a smooth
function of zero mean on $\T$.
 The special case where $V(p)=0$ (the standard Kronecker map) plays a central role here. It is known that in this case the map is uniquely ergodic
 if and only if $1,\alpha_1,\alpha_2$ are linearly independent over $\Q$. We will
 construct a quantization of it by approximating $\alpha$ with
 rational numbers $\frac{a}{N}=\frac{(a_1,a_2)}{N}$. For rational numbers we
 have an exact Egorov theorem, that is
 $$U_{a,N}^{-1}\OpN(f)U_{a,N}=\OpN(f\circ \tau_{a/N})$$
 and thus by the convergence of $\frac{a}{N}$ to $\alpha$ we
 will get (\ref{eq:egorov}). For this map we have the following theorem for polynomials:
  \begin{thm}
   Suppose $1,\alpha_1,\alpha_2$ are linearly independent over
   $\Q$. Let $f\in\Cinf(\T^2)$ be a trigonometric polynomial.
   Then for all eigenfunctions $\psi$ of $U_N(\tau_\alpha)$ we
   have that for $N$ sufficiently large
   $$\langle\OpN(f)\psi,\psi\rangle=\int_{\T^2}f(p,q)dpdq$$
  \end{thm}
 For the more general case of smooth functions we assume
 a certain restriction on $\alpha$. We assume that $\alpha$
 satisfy a certain diophantine inequality, that is there exists $\gamma>0$ such that for all
 $n_1,n_2,k\in\Z$
  \begin{equation}\label{eq:diophant}
   |n_1\alpha_1+n_2\alpha_2+k|\gg\|(n_1,n_2)\|^{-\gamma}\qquad
   (n_1,n_2)\neq (0,0)
  \end{equation}
 This reduces the set of numbers rather than being all $\alpha$
 such that $1,\alpha_1,\alpha_2$ are linearly independent over $\Q$ to
 a set of almost all $\alpha$ in Lebesgue measure sense, and $\gamma$ can be any number strictly bigger than
 2 (see theorem (\ref{Khin})\cite{S}.
 If $\alpha_1,\alpha_2$ are algebraic of degree $d_1,d_2$ respectively we can choose $\gamma$ to be $d_1!d_2!$ (\cite{S1}). For these
 $\alpha$ we have,
  \begin{thm}
   Suppose $1,\alpha_1,\alpha_2$ are linearly independent over
   $\Q$ and satisfy (\ref{eq:diophant})
   then for all $f\in\Cinf(\T^2)$ ,for all eigenfunctions $\psi$ of $U_N(\tau_\alpha)$
   $$|\langle\OpN(f)\psi,\psi\rangle-\int_{\T^2}f(p,q)dpdq|\ll N^{-\theta} \;\;\forall \theta >0$$
  \end{thm}

 Our main result is for the perturbed Kronecker map $\Phi^\alpha_V$, for arbitrary smooth $V(p)$. We show that the map is also uniquely ergodic. In fact we show that it
 is conjugate to $\tau_\alpha$ and we also have QUE for it, and
 give an upper bound for the rate of convergence:

 \begin{thm}
  Suppose $1,\alpha_1,\alpha_2$ are linearly independent over
  $\Q$ and satisfy (\ref{eq:diophant})
  then for all $f\in\Cinf(\T^2)$ ,for all eigenfunctions $\psi$ of $U_N(\Phi^\alpha_V)$
 $$|\langle\OpN(f)\psi,\psi\rangle-\int_{\T^2}f(p,q)dpdq|\ll N^{-2}$$
 \end{thm}
 Thus for such $\alpha$ the rate of convergence of the matrix elements to their classical average is much faster
 that the expected and known rates mentioned earlier on the irrational skew
 translation and the CAT map. We also construct special pairs $(\alpha_1,\alpha_2)$ and functions $f(p,q)$ for which the rate of convergence is arbitrarily slow (Theorems \ref{thm:slow_conv},\ref{thm:per_slow_conv}).
 \section*{acknowledgments}
  This work was supported in part by the EC TMR network
  \textit{Mathematical Aspects of Quantum Chaos}, EC-contract no
  HPRN-CT-2000-00103 and the Israel Science Foundation founded by
  the Israel Academy of Sciences and Humanities. This work was
  carried out as part of the author's Msc thesis at Tel Aviv
  University, under the supervision of Professor ~Zeev ~Rudnick. A
  substantial part of this work was done during the author's visit
  to the university of Bristol.

\section{Background}
 \par
 We begin with a quantization procedure for maps on the 2-torus
 $\T^2$. The procedure can be find in full description in
 \cite{KR},\cite{De}. We construct a Hilbert space of state $\mathcal{H}_h$ with respect to Planck's constant $h$, quantum
 observables, and a quantization of our maps.
  \subsection{Notations}
   We abbreviate $e(x)=e^{2\pi i x}$, and $e_N(x)=e(\frac{x}{N})$.
   $A\ll B$ or $A=O(B)$ both means that there is a constant $c$ such
   that $|A|\leq c|B|$.
  \subsection{Hilbert space of state}
   Our classical phase space is $\T^2$. The elements of the
   Hilbert space are thus, distribution on the line $\R$ that are
   periodic in both position and momentum. Using the momentum
   representation of a wave-function $\psi$ by the Fourier
   transform
   $$\mathcal{F}_h\psi(p)=\frac{1}{\sqrt{h}}\int_{-\infty}^\infty \psi(q)e(\frac{-qp}{h})dq$$
   we find that the requirements
   $$\psi(q+1)=\psi(q)\;\;\; \mathcal{F}_h\psi(p)=\mathcal{F}_h\psi(p+1)$$
   restricts planck's constant $h$ to be an inverse of integer
   $h=\frac{1}{N}$, and $\mathcal{H}_h$ consists of periodic point-mass
   distributions at the coordinates $Q=\frac{q}{N}$. We therefore
   find that the Hilbert space is of dimension $N$, and therefore
   denote $\mathcal{H}_N$, and we may identify it with
   $L^2(\Z/N\Z)$, with the inner product
   $$\langle\psi,\phi\rangle=\frac{1}{N}\sum_{Q\mod N}\psi(Q)\bar{\phi}(Q)$$
   The Fourier transform is given by
   $$\hat{\psi}(P)=\left[\mathcal{F}_N\psi\right](P)=\frac{1}{\sqrt{N}}\sum_{Q\mod N}\psi(Q)e_N(-QP)$$
   and its inverse formula is
   $$\psi(Q)=\left[\mathcal{F}_N^{-1}\hat{\psi}\right](Q)=\frac{1}{\sqrt{N}}\sum_{P\mod N}\hat{\psi}(P)e_N(PQ)$$
  \subsection{Quantum observables}
   We now assign each classical observable ,smooth functions $f\in\Cinf(\T^2$), a quantum observable, that is an operator $\OpN(f)$ on
   $\mathcal{H}_N$ that satisfy,
    \begin{enumerate}
     \item
      $\OpN(\bar{f})=\OpN(f)^*$
     \item
      $\OpN(f)\OpN(g)\sim\OpN(fg)\qquad \mbox{as } N\to \infty$
     \item
      $\frac{1}{2\pi i N}\left[\OpN(f),\OpN(g)\right]\sim\OpN(\{f,g\})\qquad \mbox{as } N\to\infty$
    \end{enumerate}
   where $\left[A,B\right]=AB-BA$ is the commutator, and
   $\{f,g\}=\frac{\partial f}{\partial p}\frac{ \partial g}{\partial q}-\frac{\partial g}{\partial p}\frac{\partial p}{\partial q}$
   are the Poisson bracket.
   The norm used is the induced norm from the inner
   product on $\mathcal{H}_N$.
   \par
   The translation operators
   $$\left[{\rm t}_1\psi\right](Q)=\psi(Q+1)$$
   and
   $$\left[{\rm t}_2\psi\right](Q)= e_N(Q)\psi(Q)$$
   play a special role, they are analogues of the of the differentiation and
   multiplication operators. ~Heisenberg's
   commutation relations are
   $${\rm t}_1^a{\rm t}_2^b={\rm t}_2^b{\rm t}_1^a e_N(ab)\qquad \forall a,b\in\Z$$
   Notice that
   $$\mathcal{F}_N{\rm t}_1\mathcal{F}_N={\rm t}_2$$
   and
   $$\mathcal{F}_N{\rm t}_2\mathcal{F}_N={\rm t}_1^{-1}$$
   With these operators we construct
   $$T_N(n)=e_N(\frac{n_1 n_2}{2})t_2^{n_2}t_1^{n_1}, n=(n_1,n_2)\in \Z^2$$
   whose action on a wave-function $\psi\in\mathcal{H}_N$ is
   $$T_N(n)\psi(Q)=e^{\frac{i\pi n_1 n_2}{N}}e_N(n_2 Q)\psi(Q+n_1)$$
   Notice that
   $$T_N(n)^*=T_N(-n)$$
    \begin{equation}\label{omega}
     T_{N}(m)T_{N}(n)=e_{N}({\frac{\omega(m,n)}{2}})T_{N}(m+n)
    \end{equation}
   where, $\omega(m,n)=m_{1}n_{2}-m_{2}n_{1}$, and that $T_N$ is a
   unitary operator.
   Finally for a general smooth function
   $$f(x)=\sum_{n\in\Z^2}\hat{f}(n)e(n\cdot x)$$
   where $x=(p,q)$.
   we define its quantization $\OpN(f)$
    \begin{equation} \label{no1}
     \OpN(f)= \sum_{n\in Z^2} \hat{f}(n)T_N(n)
    \end{equation}
   and the conditions mentioned are all satisfied.
\section{Quantization of maps and rate of convergence}
 When quantizing a map, we look for a sequence of unitary operators $U_N(A)$ on
 $\mathcal{H}_N$, the {\it quantum propagator}, whose iterates
 give the evolution of the quantum system, and that in the semiclassical limit, (the limit as $N\to\infty$ or
 $h\to\infty$), the quantum evolution follows the classical
 evolution as described in the following definition.
 \begin{defn}["Egorov's Theorem"]
  A quantization of a continuous map $A:\T^2\to\T^2$ is a sequence of unitary
  operators ,$\{U_N\}$, satisfying:
   \begin{equation}\label{quant}
    \|{U_{N}^{-1}\OpN(f)U_{N}-\OpN(f\circ A)}\|\to0 \qquad
    {\mbox{as }} N\to\infty
   \end{equation}
 \end{defn}

 The stationary states of the quantum system are given by the
 eigenfunctions $\psi$ of $U_N(A)$. We will find that for the maps studied in this paper the limiting
  expectation value of observables in normalized eigenstates
  converges to the classical average of the
  observable, that is
  $$\langle\OpN(f)\psi,\psi\rangle\to\int_{\T^2}f\qquad {\mbox{as  }} N\to\infty$$

\subsection{Quantizing ~Kronecker map }
 In this section we will construct a quantization of the ~Kronecker map.
  \begin{eqnarray*}
   \tau_\alpha:& \T^2 \to    &\T^2\\
               &
                    \begin{pmatrix}
                     p \\
                     q
                    \end{pmatrix}
                   \mapsto &
                    \begin{pmatrix}
                     p+\alpha_1 \\
                     q+\alpha_2+V(p)
                    \end{pmatrix} \mod 1
  \end{eqnarray*}
 \begin{lem}\label{lem:egorov_for_Kron}
  suppose $\frac{(a_1,a_2)}{N}$ is a sequence of rational numbers
  such that
  $\frac{(a_1,a_2)}{N}=\frac{\vec{a}}{N}_{\overrightarrow{N\to\infty}}\vec{\alpha}$
  then the sequence $U_N(\tau_\alpha):=T_N(-a_2,a_1)$ is a quantization of
  Kronecker's map.
 \end{lem}
 \begin{proof}
  First assume $f(x)=e_n(z):=e(n\cdot z)$ in this case we get
  $\hat{f}(n)=1,\hat{f}(m)=0$ for $m\neq n$ ,and therefore $\OpN(f)=T_N(n)$.\\
  Denote $\tilde{a}:=(-a_{2},a_{1})$,and notice that  $n\cdot
  a=\omega(n,\tilde{a})$.  Now
  $$U_{N}(\tau_\alpha)^{-1}T_{N}(n)U_{N}(\tau_\alpha)=T_{N}(-\tilde{a})T_{N}(n)T_{N}(\tilde{a})$$
  which due to (\ref{omega}) and the linearity and antisymmetry of $\omega(m,n)$
   \begin{equation}\label{RHS}
    e_N(\omega(n,\tilde{a}))T_{N}(n)=e_N(n\cdot a))T_{N}(n)
   \end{equation}
  on the other hand,we have
   \begin{equation*}
    (e_n\circ\tau_{\alpha})(x)=e(n_1(p+\alpha_1)+n_2(q+\alpha_2))=e(n\cdot\vec{\alpha})e_n(x)
   \end{equation*}
  and so
   \begin{equation}\label{LHS}
    \OpN(e_n\circ \tau_{\alpha})=e(n\cdot\vec{\alpha})T_{N}(n)
   \end{equation}
  From (\ref{RHS}),(\ref{LHS}) we get that
  $$\|U_{N}^{-1}(\tau_\alpha)T_N(n)U_{N}(\tau_\alpha)-e(n\cdot\alpha)T_{N}(n)\|=|e_{N}(n\cdot\vec{a})-e_{N}(n\cdot\vec{\alpha})|\cdot\|T_{N}(n)\|$$
  $T_N$ is a unitary operator so $\|T_N(n)\|=1$ we get
  $$|e_{N}(n\cdot{\frac{\vec{a}}{N}})-e_{N}(n\cdot\vec{\alpha})|\ll \|n\| |\vec{\alpha}-\frac{\vec{a}}N|$$
  Therefore we established (\ref{quant}) for $f=e_n(x)$. By
  linearity we also have (\ref{quant}) for trigonometric
  polynomials. suppose now that $f(x)$ is a general function of
  $C^\infty(\mathbb{T}^2)$ and therefore
  $$f(x)=\sum_{n\in\mathbb{Z}^2}\hat{f}(n)e_n(x)$$
  Consider
   \begin{multline*}
    \|U_{N}^{-1}(\tau_\alpha)\OpN(f)U_{N}(\tau_\alpha)-\OpN(f\circ A)\|=\\
    \|U_{N}^{-1}(\tau_\alpha)\{\sum_{n\in\mathbb{Z}^2}\hat{f}(n)T_N(n)\}U_N(\tau_\alpha)-\sum_{n\in\mathbb{Z}^2}\hat{f}(n)e(n\cdot\alpha)T_{N}(n)\|=\\
    \|\sum_{n\in\mathbb{Z}^2}\hat{f}(n)\{e_N(n\cdot a)-e(n\cdot\alpha)\}T_N(n)\|\leq
    \sum_{n\in\mathbb{Z}^2}|\hat{f}(n)|\cdot |e(n\cdot a)-e(n\cdot\alpha)|\cdot\| T_N(n)\|
   \end{multline*}
  and therefore
   \begin{eqnarray*}
    \|U_{N}^{-1}(\tau_\alpha)\OpN(f)U_{N}(\tau_\alpha)-\OpN(f\circ A)\|=|\vec{\alpha}-\frac{\vec{a}}N|\sum_{n\in\mathbb{Z}^2}\|n\|\hat{f}(n)=O(|\vec{\alpha}-\frac{\vec{a}}N|)
   \end{eqnarray*}
  which goes to zero since $|\vec{\alpha}-\frac{\vec{a}}N|\to0$ as $N\to\infty$
  implying that $U_N$ is a quantization of $\tau_\alpha$.\\
  \begin{remark}
   Notice that for each $N$, we have exact Egorov for $\tau_{a/N}$,
   that is
   $$U_{N}^{-1}(\tau_{a/N})\OpN(f)U_{N}(\tau_{a/N})=\OpN(f\circ\tau_{a/N})$$
  \end{remark}
 \end{proof}
\subsection{Convergence of eigenstates}\label{sec:ass behav}
 We now wish to give an upper bound for the remainder
  \begin{equation}\label{eq:shnir}
   |\langle\OpN(f)\psi,\psi\rangle-\int_{T^2}f|
  \end{equation}
 where $\psi$ is an eigenfunction of $U_N$.
 Actually we will prove the following two theorems:
 \begin{thm}\label{thm:assbehav}
  Suppose $1,\alpha_1,\alpha_2$ are linearly independent over $\mathbb{Q}$. Then
  For any eigenfunction $\psi(Q)$ of $U_N$
  \begin{enumerate}
   \item
    If $f$ is a polynomial then for $N$ large enough,
    $$\langle\OpN(f)\psi,\psi\rangle=\int_{T^2}f$$
   \item
    if $\alpha=(\alpha_1,\alpha_2)$ is diophantine (see definition \ref{def:dioph}) and $|\vec{\alpha}-\frac{\vec{a}}N|\ll\frac{1}{N}$
    then for all $f\in\Cinf(\T^2)$
    $$\langle\OpN(f)\psi,\psi\rangle-\int_{T^2}f= O(\frac{1}{N^\theta}) \;\;,\forall\theta>0$$
  \end{enumerate}
 \end{thm}
 \begin{thm}\label{thm:slow_conv}
  For any positive increasing function $g(x)$, there exists $\alpha=(\alpha_1,\alpha_2)$ such that $1,\alpha_1,\alpha_2$ are linearly independent over the rationals ,$f\in\Cinf(\T^2)$, and a basis of eigenfunctions $\{\psi_j\}_{j=1}^N$ such
  that
  $$|\langle\OpN(f)\psi_j,\psi_j\rangle-\int_{T^2}f|\gg\frac{1}{g(N)}$$
 \end{thm}
 \begin{remark}
  The set of all diophantine pairs is of Lebesgue  measure 1 (see theorem
  \ref{Khin}). An example for such pairs are $\alpha=(\alpha_1,\alpha_2)$ such that
  $\alpha_1,\alpha_2$ are algebraic and $1,\alpha_1,\alpha_2$ are
  linearly independent over $\Q$ (see theorem \ref{thm:alg_dioph}).
 \end{remark}
 To prove these theorems we will start with the following lemma:
 \begin{lem}\label{lem:mat_elem}
  Let $\psi(Q)$ to be an eigenfunctions of $U_N$.
  \begin{enumerate}
   \item
    \begin{equation}\label{eq:time_average}
     \langle\OpN(f)\psi,\psi\rangle=\langle\OpN(f^T)\psi,\psi\rangle
    \end{equation}
    where
    $$f^T(p,q)=\frac{1}{T}\sum_{t=0}^{T-1}f\circ\tau_{(a/N)}^t$$
   \item
    For $f(x)=e_n(x)$, $\langle \OpN(f)\psi,\psi\rangle$ is identically zero for large enough N.
  \end{enumerate}
 \end{lem}
 \begin{proof}
  \begin{enumerate}
   \item
    Since $\psi$ is an eigenfunction of $U_N$ then
    $U_N\psi=e(\phi)\psi$, and therefore for all $t$
    $$\langle\OpN(f)U_N^t\psi,U_N^t\psi\rangle=\langle e(t\phi)\OpN(f)\psi,e(t\phi)\psi\rangle=\langle\OpN(f)\psi,\psi\rangle$$
    Now,
    $$\langle\OpN(f)U_N^t\psi,U_N^t\psi\rangle=\langle U_N^{-t}\OpN(f)U_N^t\psi,\psi\rangle$$
    and since
    $$U_N^{-t}\OpN(f)U_N^t=\OpN(f\circ\tau_{a/N}^t)$$
    we have (\ref{eq:time_average}).
   \item
    fix $\vec{n}=(n_{1},n_{2})\in \Z^{2}$ , $f(x)=e_n(x)$ and therefore \\
    $\OpN(f)=T_{N}(n)$. Notice that for  $f=e_n$ we have,
     \begin{eqnarray*}
      f^T=\frac{1}{T}\sum_{t=0}^{T-1}e_n\circ\tau_{(a/N)}^t=\frac{1}{T}\sum_{t=0}^{T-1}e(n_1(p+t a_1/N)+n_2(q+t a_2/N))=\\
      \frac{1}{T}e_n(p,q)\sum_{t=0}^{T-1}e_N((n_1a_1+n_2a_2)t)
     \end{eqnarray*}
    and for $T=N$ we have,
     \begin{equation}
      f^N=\begin{cases}
           f &{\rm if}\; n_2a_2+n_1a_1=0\pmod N\\
           0 &{\rm else}
          \end{cases}
     \end{equation}
    and therefore,
     \begin{equation}\label{eq:most_elem_zero}
      \OpN(f^N)=\begin{cases}
                 \OpN(f) &{\rm if}\; n_2 a_2+n_1a_1=0\pmod N\\
                 0 &{\rm else}
                \end{cases}
     \end{equation}

    but
     \begin{eqnarray*}
      n_2a_2+n_1a_1=Nk \Longleftrightarrow
      n_2\frac{a_2}{N}+n_1\frac{a_1}{N}=k&\in&\Z\\
      \Longleftrightarrow  n_2\{\alpha_{2}+{\rm O}(|\vec{\alpha}-\frac{\vec{a}}N|)\}+n_1\{\alpha_{1}+{\rm O}(|\vec{\alpha}-\frac{\vec{a}}N|)\}=k&\in &\Z
     \end{eqnarray*}
    and so we get
     \begin{equation}\label{eq:alpha}
      n_2\alpha_2+n_1\alpha_1+{\rm O}(\|n\||\vec{\alpha}-\frac{\vec{a}}N|)\}=k\in \Z
     \end{equation}
      $\alpha_1,\alpha_2$ are linearly independent over $\mathbb{Q}$ so
    we can denote $0<\delta=dist(n_1\alpha_1+n_2\alpha_2,\mathbb{Z})$.
    Now assume that there exists infinitely many pairs
    $\vec{a}=(a_1,a_2)$ such that (\ref{eq:shnir}) is nonzero i.e.
    $n_2a_2+n_1a_1=Nk_{\vec{a}}$ .From (\ref{eq:alpha})
    we get that
     \begin{equation}
      {\rm O}(\|n\||\vec{\alpha}-\frac{\vec{a}}N|)=|k+n_{2}\alpha_2+n_1\alpha_1|\geqslant\delta>0,N\to\infty
     \end{equation}
    now since n is fixed and $|\vec{\alpha}-\frac{\vec{a}}N|\to 0$ as $N\to\infty$ we get a contradiction! so we can deduce that for $N\gg\|n\|$
    $$|\langle \OpN(f)\psi,\psi\rangle-\int_{\mathbb{T}^2}f|^{2}=|\langle T_N(n)\psi,\psi\rangle|=0$$
   \end{enumerate}
 \end{proof}
 \begin{cor}[QUE for Kronecker map]\label{thm:QUE_Kron}
  For any eigenfunction $\psi$ of $U_N$,
  \begin{enumerate}
   \item
    if $f$ is a trigonometric polynomial,
    $\langle\OpN(f)\psi,\psi\rangle=\int_{\T^2}$ for large enough N.
   \item
    For any $f\in\Cinf(\T^2)$,
    $$|\langle\OpN(f)\psi,\psi\rangle-\int_{T^2}f|\to0 \qquad \mbox{as } N\to\infty$$
  \end{enumerate}
 \end{cor}
 \begin{proof}
  \begin{enumerate}
   \item
    From the previous lemma we get that every trigonometric function
    has N such that ({\rm \ref{eq:shnir}}) is identically zero so for
    a finite linear combination
    $$\sum_{n=1}^m a_n e(n\cdot x)$$
    simply choose the largest N given from $e_n(x),n=1,\dots,m$
   \item
    For a general $f\in\Cinf(\T^2)$, we have
    $$\OpN(f)=\sum_{n\in\Z^2}\hat{f}(n)T_N(n)$$
    For $\epsilon>o$,
    there exists $R_0$, such that $\forall R>R_0$,
    $$\sum_{\|n\|>R}|\hat{f}(n)|<\epsilon$$
    For the polynomial
    $$P_R=\sum_{\|n\|<R}\hat{f}(n)e(n\cdot x)$$
    there exists $N_0$, such that for all $N>N_0$
    $$\langle\OpN(P_R)\psi,\psi\rangle=0$$
    and so we have ,
    \begin{multline*}
    |\langle\OpN(f)\psi,\psi\rangle|\leqslant\\
    |\langle\OpN(P_R)\psi,\psi\rangle|+|\sum_{\|n\|>R}\hat{f}(n)\langle T_N(n)\psi,\psi\rangle|\leqslant \epsilon
    \end{multline*}
    for $N>N_0$.
  \end{enumerate}
 \end{proof}
 \subsubsection{Convergence of eigenstates for diophantine pairs:}
 To finish the study of the upper bound for a general
 function we need to study the size of $n_1\alpha_1+n_2\alpha_2+k$
 for $n_1,n_2,k\in\Z$,and assume that $\alpha$ satisfies a certain
 diophantine inequality that is
 $|n_1\alpha_1+n_2\alpha_2+k|\gg \frac{c(\alpha)}{\|n\|^{\gamma}}$ for some
 $\gamma$. Numbers like this are called diophantine.
  \begin{defn}\label{def:dioph}
   An l-tuple of real numbers $(\alpha_1,\dots,\alpha_l)$ is called
   diophantine if they satisfy that there exists $\gamma$ such that for any
   integers $(n_1,\dots,n_l)\neq \vec{0},k$
   $$|n_1\alpha_1+\cdots n_l\alpha_l+k|\gg\frac{c(\alpha)}{\|n\|^\gamma}$$
  \end{defn}
 with this we have the following.
 \begin{cor}\label{cor:alg}
  Suppose $\alpha$ is diophantine and that
  $|\vec{\alpha}-\frac{\vec{a}}N|\ll\frac{1}{N}$ then
  we have an upper bound for $|\langle\OpN(f)\psi,\psi\rangle-\int_{\T^2}f|\ll\frac{1}{N^\theta}$ for any $\theta>0$.
 \end{cor}
 \begin{proof}
  A general function is of the following form
  $$f(x)=\sum_{n\in\Z^2}\hat{f}(n)e_n(x)$$
  without loss of generality we can assume that
  $\int_{\mathbb{T}^2}f=0$ and so divide $\OpN (f)$ into two sums:
  $\OpN(f)=\sum_{n\in\Z^2}\hat{f}(n)T_N(n)=I_1+I_2$ where
  $I_1=\sum_{\|n\|\leqslant R}\hat{f}(n)T_N(n),I_2=\sum_{\|n\|>R}\hat{f}(n)T_N(n)$.
  Now as seen earlier, the case when $|\langle T_{N}(\vec{n})\psi,\psi\rangle|\neq 0$
  can only happen when
  $${\rm O}(\frac{\|n\|}{N})=k+n_{2}\alpha_2+n_1\alpha_1$$
  but our assumption is that there exists $\gamma$
  such that for all integer coefficients
  $k+n_{2}\alpha_2+n_1\alpha_1\gg\frac{1}{\|n\|^{\gamma}}\gg\frac{1}{R^{\gamma}}$
  and so define $N=R^{1+\gamma+\delta}$ for some
  $\delta>0$ and we get that
   \begin{equation*}
    \frac{R}{N}\geqslant\frac{\|n\|}{N}\gg k+n_2\alpha_2+n_1\alpha_1\gg\frac{1}{\|n\|^\gamma}\gg\frac{1}{R^\gamma}
   \end{equation*}
  and for $N=R^{1+\gamma+\delta}$ this gives a
  contradiction and so $I_1=0$ for large enough N.
  For $I_2$ we use the rapid decay of the Fourier coefficients:
  $$|I_2|=|\sum_{\|n\|>R}\hat{f}(n)T_N(n)|\leqslant\sum_{\|n\|>R}\|\hat{f}(n)T_N(n)\|=
  \sum_{\|n\|>R}|\hat{f}(n)|\leqslant\frac{1}{R^b}=\frac{1}{N^\theta}$$
  for any chosen $\theta$.
 \end{proof}
 For algebraic numbers we have this inequality by the following
 well known theorem, (\cite{S1}):
 \begin{thm}\label{thm:alg_dioph}
  Suppose $\vec{\alpha}=(\alpha_1,\dots,\alpha_m)$ are linearly
  independent over $\mathbb{Q}$ then there exists $D=D(\alpha)$ such
  that
  $$|n_1\alpha_1+n_m\alpha_m+k|\gg\frac{c(\vec{\alpha})}{\|n\|^{D-1}}$$
 \end{thm}
 For the more general $\vec{\alpha}$ we need the following theorem
 by Khintchine \cite{S}:
 \begin{thm}\label{Khin}
  Almost no pair $(\alpha_1,\alpha_2)$ is very well approximable
  that is that for almost any pair there exists
  $\delta=\delta(\alpha_1,\alpha_2)$ such that there are only finite
  many integers $m=(m_1,m_2),k$ such that the following inequality
  holds:$|m_1\alpha_1+m_2\alpha_2+k|\geqslant\frac{1}{\|m\|^{2+\delta}}$
 \end{thm}
\subsubsection{Proof of theorem \ref{thm:slow_conv}}
 We begin the proof using a construction of an irrational number
 $\alpha$, and a sequence converging to it.
 \begin{lem}\label{lem:irrat_const}
 Given any positive increasing function $F(x)$ there is an
irrational $\beta$ with continued fraction expansion
$[b_1,b_2,\dots,b_n,\dots]$, such that the partial quotients
$c_n/d_n=[b_1,\dots,b_n]$ satisfy:
\begin{enumerate}
\item
$F(d_n)\leq b_{n+1}d_n^2$
\item
$|\beta-\frac{c_n}{d_n}|<\frac{1}{F(d_n)}$
\end{enumerate}
\end{lem}
The proof of the lemma is given in \cite{MR}.
 Set $G(x)=\log g(x)$, and apply lemma \ref{lem:irrat_const} for $F=G^{-1}$.
 Following the lemma's notation define $f(p,q)=\sum_{n=1}^\infty e^{-d_n}e(d_n q)$,
$\alpha=(\sqrt{2},\beta),b=b_{n+1}c_nd_n,N=b_{n+1}d_n^2$
\begin{thm}
For $\alpha,f(p,q),b,N$ defined above the following holds:
\begin{enumerate}
\item
 $U_N=T_N(-b,a)$ is a quantization of $\tau_\alpha$, where
 $\frac{a}{N}$ is a sequence converging to $\sqrt{2}$.
\item
 There exists a basis of eigenfunctions $\{\psi_j\}_{j=1}^N$ of $U_N$ such that
 \begin{equation}\label{eq:spec_mat_elem}
 |\langle T_N(0,d_n)\psi_j,\psi_j\rangle|=1
 \end{equation}
\item
 For the basis $\{\psi_j\}_{j=1}^N$
$$|\langle\OpN(f)\psi_j,\psi_j\rangle|\gg\frac{1}{g(N)}$$
\end{enumerate}
\end{thm}
\begin{proof}
\begin{enumerate}
\item
 According to the construction from lemma \ref{lem:irrat_const} we
 get that $|\beta-\frac{b}{N}|\to0$ as $N\to\infty$, and therefore
 $\frac{(a,b)}{N}$ converges to $\alpha$, and thus by lemma
 \ref{lem:egorov_for_Kron} we have that $U_N$ is a quantization of
 $\tau_\alpha$.
\item
 Since $\omega((0,d_n),(-b,a))=d_nb=c_nN\equiv0\pmod N$ we have
 that $T_N(0,d_n),T_N(-b,a)$ commute (according to (\ref{omega})),
 and therefore they have an orthonormal basis of joint eigenfunctions
 $\{\psi_j\}_{j=1}^N$, and since $T_N(n)$ is a unitary operator we
 have
$$ |\langle T_N(0,d_n)\psi_j,\psi_j\rangle|=|e(\phi)\langle
\psi_j,\psi_j\rangle|=1$$ as required
\item
 We first observe that
 \begin{equation}\label{eq:fourier_coeff}
\hat{f}(n_1,n_2)= \begin{cases}
                    e^{-d_n} & (n_1,n_2)=(0,d_n)\\
                    0        & \mbox{otherwise}
                   \end{cases}
\end{equation}
By definition of $\OpN(f)$ we have that
\begin{equation}\label{eq:mat_elem}
\langle\OpN(f)\psi,\psi\rangle=\sum_{n\in\Z^2}\hat{f}(n)\langle
T_N(n)\psi,\psi\rangle
\end{equation}
From equation (\ref{eq:most_elem_zero}) we saw that
$$n_1a+n_2b\not\equiv 0\pmod N\Rightarrow \langle T_N(n)\psi,\psi\rangle=0$$
and therefore the RHS in (\ref{eq:mat_elem}) is in fact
$$\sum_{n_1a+n_2b\equiv 0\pmod N}\hat{f}(n)\langle T_N(n)\psi,\psi\rangle$$
Form (\ref{eq:fourier_coeff}) we see that if $n_1\neq0$ then
$\hat{f}(n_1,\cdot)=0$ and thus the condition
$n_1a+n_2b\equiv0\pmod N$ is in fact $n_2b\equiv0\pmod
N\Leftrightarrow n_2\equiv0\pmod {\frac{N}{(b,N)}}\Leftrightarrow
n_2\equiv0\pmod {d_n}$ (the last equality is by definition of
$b,N$), and therefore
\begin{eqnarray*}
\langle\OpN(f)\psi_j,\psi_j\rangle=\sum_{n_2\equiv0\pmod{d_n}}\hat{f}(0,n_2)\langle
T_N(0,n_2)\psi_j,\psi_j\rangle\geq\\
|\hat{f}(0,d_n)\langle
T_N(0,d_n)\psi_j,\psi_j\rangle|-\sum_{k=2}^\infty|\hat{f}(0,kd_n)\langle
T_N(0,kd_n)\psi_j,\psi_j\rangle|\geq\\
|\langle
T_N(0,d_n)\psi_j,\psi_j\rangle|e^{-d_n}-\frac{e^{-2d_n}}{1-e^{-d_n}}=\\
e^{-d_n}|\langle
T_N(0,d_n)\psi_j,\psi_j\rangle-\frac{e^{-2d_n}}{1-e^{-d_n}}|\geq
e^{-d_n}|1-\frac{e^{-2d_n}}{1-e^{-d_n}}| \gg e^{-d_n}
\end{eqnarray*}
and since $F(d_n)\leq N$ we have that $d_n\leq G(N)=\log g(N)$ and
therefore $e^{-d_n}\geq e^{-\log g(N)}=\frac{1}{g(N)}$ and we get
that
$$\langle\OpN(f)\psi_j,\psi_j\rangle\gg\frac{1}{g(N)}$$

\end{enumerate}
\end{proof}

\subsection{Perturbed Kronecker map}
 Another family of uniquely ergodic maps on $\T^2$, is the
 perturbed Kronecker map. we see in this section that it is
 uniquely ergodic, due to the fact that it is conjugate to the
 Kronecker map itself, and in the following section we form a
 quantization for it.
 Define the following shear perturbation:
 $$\Phi_V : \begin{pmatrix}
             p \\
             q
            \end{pmatrix}
   \mapsto
   \begin{pmatrix}
    p \\
    q+V(p)
   \end{pmatrix}$$
 and the perturbed Kronecker map:
 $$\Phi^\alpha_V :
   \begin{pmatrix}
    p \\
    q
   \end{pmatrix}
   \mapsto
    \begin{pmatrix}
     p+\alpha_1 \\
     q+\alpha_2+V(p)
    \end{pmatrix}$$
 where $V(p)\in\Cinf(\T)$ satisfies $\int_0^1 V(p)dp=0$. In order to
 prove the unique ergodicity of this map, we will use the following
 Lemma that shows that the perturbed map is conjugate to the Kronecker map.
 \begin{lem}\label{lem:pert_conj}
  Suppose $\alpha_1$ is irrational.
   \begin{enumerate}
    \item If $V(p)$ is a polynomial we have that
     \begin{equation*}\label{eq:pert_conj}
      \tau_\alpha\circ\Phi_{V}=\Phi_{h}\circ\tau_\alpha\circ\Phi_{h}^{-1}
     \end{equation*}
    \item If $\alpha_1$ is diophantine then (\ref{eq:pert_conj}) holds for any
     $V\in\Cinf(\T)$
   \end{enumerate}
  for some $h=h_V\in\Cinf(\T)$
 \end{lem}
 \begin{proof}
  \begin{enumerate}
   \item The RHS of (\ref{eq:pert_conj}) is
    \begin{equation*}
     \Phi_{h_k}\circ\tau_\alpha\circ\Phi_{h_k}^{-1}(p,q)=
     \begin{pmatrix}
      p+\alpha_1\\
      q+\alpha_2+h_k(p+\alpha_1)-h_k(p)
     \end{pmatrix}
    \end{equation*}
    define $h_k(p)=\frac{e(kx)}{e(k\alpha_1)-1}$ (which is well defined for all $k$ only if $\alpha_1$ is irrational). $h_k(p)$ satisfy
    that $e(kp)=h_k(p+\alpha_1)-h_k(p)$ and therefore we get
    (\ref{eq:pert_conj}),and by linearity we get that(\ref{eq:pert_conj}) holds for every polynomial.
   \item For $V\in\Cinf(\T)$, $\alpha$ diophantine , we observe that
    $|e(k\alpha_1)-1| \sim \{k\alpha\}\gg\frac{1}{|k|^\gamma}$ and
    we get that
    $$\sum_{k\in\Z}|\hat{V}(k)h_k(p)|\ll\sum_{k\in\Z}|\hat{V}(k)||k|^{\gamma}$$
    converges absolutely and so define
    $h_V(p)=\sum_{k\in\Z}\hat{V}(k)h_k(p)$. Then $h_V(p)$
    satisfies $h_V(p+\alpha_1)-h_V(p)=V(p)$ since $h_k$ satisfy that for every
    $k$ since the series converges absolutely.
  \end{enumerate}
 \end{proof}
 With $\Phi^{\alpha}_V$ described as a conjugate of $\tau_\alpha$
 we have the following result:
 \begin{thm}
  Suppose $1,\alpha_1,\alpha_2$ are linearly independent over
  $\mathbb{Q}$. Then for $\alpha$ diophantine and $V(p)\in\Cinf(\T)$ then
     $\Phi^{\alpha}_V$ is uniquely ergodic.
 \end{thm}
 \begin{proof}
  We will first show that Lebesgue measure is $\Phi^{\alpha}_V$
  invariant. Suppose $f(p,q)\in L^1(\T^2)$. Then
  $f\circ\Phi^{\alpha}_V(p,q)=f(p,q+V(p))$ and so
   \begin{equation*}
    \int_0^1\int_0^1 f(p+\alpha_1,q+V(p)+\alpha_2)dqdp=\int_0^1\int_0^1 f(p,q)dqdp
   \end{equation*}
  by standard change of variables.
  Now, assume $\mu$ is an invariant measure of $\Phi^{\alpha}_V$. since
  $\Phi^{\alpha}_V=\Phi_{h}\circ\tau_\alpha\circ\Phi_{h}^{-1}$ for
  some $h\in\Cinf(\T)$, then $\Phi_h\circ\mu$ is invariant measure of
  $\tau_\alpha$, but there exists only one such measure and which
  is Lebesgue measure $m$, that is $\Phi_h\circ\mu=m$ is Lebesgue
  measure. $\Phi_h$ is an invertible map, that preserves Lebesgue
  measure, so $\mu=\Phi^{-1}_h\circ m=m$
  therefore $\Phi^{\alpha}_V$ is uniquely ergodic.
 \end{proof}
\subsection{QUE for perturbed  Kronecker map}
 In this section we will study the asymptotic behaviour of the matrix elements related
 to the perturbed Kronecker map. The main tool will be
 lemma \ref{lem:pert_conj} that connects the perturbed map to the unperturbed map.

In order to quantize the perturbed Kronecker map, we use the
following theorem of Marklof-O'Keefe \cite{MO}:
 \begin{thm}[Marklof-O'Keefe]\label{egorov for pert}
  For every function $f\in \Cinf(\T^2)$ we have
   \begin{equation}
    |\langle \left( U_v^{-1}\OpN(f)U_v -\OpN(f\circ \Phi_v)\right)\psi,\psi \rangle|\ll\frac{c(f)}{N^2}
   \end{equation}
 \end{thm}

 Using the equality in Lemma (\ref{lem:pert_conj}) and the quantization of the perturbation map in theorem \ref{egorov for pert}, we can describe
 the quantization of $\Phi_V^\alpha=\tau_\alpha\circ\Phi_v$ as follows:
 \begin{thm}
  Denote $U_N=U_h(N)^{-1}U_{\tau}(N)U_h(N)$ where $U_{\tau}(N)$ is the
  quantization of $\tau_\alpha$, then we have
   \begin{equation}
    \|U_N^{-1}\OpN(f)U_N-\OpN(f\circ\Phi_v^\alpha)\|\ll N^{-1}
   \end{equation}
 \end{thm}
 \begin{proof}
  We already know that
   \begin{equation*}
    \|U_h^{-1}\OpN(f)U_h-\OpN(f\circ\Phi_h)\|=O(N^{-2})
   \end{equation*}
  and that
   \begin{equation*}
    \|U_{\tau}(N)^{-1}\OpN(f)U_{\tau}-\OpN(f\circ\tau)\|=O(N^{-1})
   \end{equation*}
  and thus using the equality in Lemma (\ref{lem:pert_conj}) we conclude
  the proof
 \end{proof}
 \begin{remark}
  The set $\{\psi_j=U_h(N)^{-1}\psi^{\tau}_j\}$ form a basis of
  eigenfunctions of $U_N$ , where $\{\psi^{\tau}_j\}$ is a basis
  of eigenfunctions for $U_{\tau}$.
 \end{remark}
 With this representation of the eigenfunctions we can give an
 upper bound for the asymptotic behavior of the matrix elements:
 \begin{thm}
  For every $f\in\Cinf(\mathbb{T}^2)$, $\alpha$ diophantine we
  have:
 $$|\langle\OpN(f)\psi_j,\psi_j\rangle-\int f|\ll N^{-2}$$
 \end{thm}
 \begin{proof}
  Without loss of generality we will assume that $\int f=0$. By
  definition we have
  $$\langle\OpN(f)\psi_j,\psi_j\rangle=\langle\OpN(f)U_h^{-1}\psi^{\tau}_j,U_h^{-1}\psi^{\tau}_j\rangle$$
  and since $U_h$ is unitary we have
  $$\langle\OpN(f)\psi_j,\psi_j\rangle=\langle U_h\OpN(f)U_h^{-1}\psi^{\tau}_j,\psi^{\tau}_j\rangle$$
  Now using Theorem \ref{egorov for pert} we get,
  \begin{equation}\label{eq:conj_rate}
  |\langle U_h\OpN(f)U_h^{-1}\psi^{\tau}_j,\psi_j^\tau\rangle-\langle \OpN(f\circ\Phi_h)\psi^{\tau}_j,\psi_j^\tau\rangle|\ll N^{-2}
  \end{equation}
  since $\psi_j$ is a normalized wavefunction , but since
  $f\circ\Phi_h$ is still a $\Cinf(\T^2)$ we have that
  the second term is $O(N^{-10})$ and therefore
  $$\langle \OpN(f)\psi_j,\psi_j\rangle\ll N^{-2}$$
 \end{proof}
 \begin{remark}
  The upper bound found here is valid only for the quantization of
  described here which includes an arbitrary choice of a sequence
  that converges to $\alpha$ by rational numbers. Since this
  quantization is not unique, and since the operators
  $\|U_N(a)-U_N(a')\|\sim\frac{1}N$ this upper bound only applies
  with the specific eigenfunctions for a specific chosen convergent
  sequence for $\alpha$.
 \end{remark}
As for the standard ~Kronecker map , we can also construct special
$\alpha,f\in\Cinf(\T^2)$ with arbitrary slow convergence:
\begin{thm}\label{thm:per_slow_conv}
For any positive increasing $g(x)$ there exist
$\alpha,\tilde{f}(p,q)\in\Cinf(\T^2)$ and a basis of
eigenfunctions $\{\psi_j\}_{j=1}^N$ such that
$$\langle
\OpN(\tilde{f})\psi_j,\psi_j\rangle-\int_{\T^2}\tilde{f}|\gg\frac{1}{g(N)}$$
\end{thm}
\begin{proof}
Take $\alpha$ to be the pair $(\sqrt{2},\beta)$ as in theorem
\ref{thm:slow_conv}. Since $\sqrt{2}$ is diophantine then
$\Phi_V^\alpha$ is still conjugate to $\tau_\alpha$, we still have
$$|\langle U_h\OpN(\tilde{f})U_h^{-1}\psi^{\tau}_j,\psi_j^\tau\rangle-\langle \OpN(\tilde{f}\circ\Phi_h)\psi^{\tau}_j,\psi_j^\tau\rangle|\ll N^{-2}$$
And thus for $\tilde{f}=f\circ\Phi_h^{-1},\{\psi^{\tau}_j\}$ where
$f(p,q),\psi^{\tau}_j$ are the function and orthonormal basis
constructed for the proof of theorem \ref{thm:slow_conv} we have
$$|\langle\OpN(\tilde{f})\psi_j,\psi_j\rangle-\int_{\T^2}\tilde{f}|\gg\frac{1}{g(N)}$$
\end{proof}
\begin{remark}
Notice that due to corollary \ref{thm:QUE_Kron}, and
(\ref{eq:conj_rate}), the matrix elements do converge to
$\int_{\T^2}f$ and thus we still have QUE, but with rate of
convergence arbitrary slow
\end{remark}


\begin{thebibliography}{10}

\bibitem{C} Y. Colin de Verdiere, Ergodicit´e et functions propres du Laplacian, {\em Comm. Math. Phys.} {\bf 102} (1985), 497-502.

\bibitem{De} M. Degli Esposti, Quantization of the orientation
preserving automorphisms of the torus, {\em Ann. Inst. Poincar\'e}
{\bf 58} (1993) 323-341.

\bibitem{DEGI} M. Degli Esposti, S. Graffi and S. Isola {\em Classical
 limit of the quantized hyperbolic toral automorphisms}, Comm. Math
 Phys. {\bf 167} (1995), 471--507.


\bibitem{DebFaNon}  F.\,Faure, S.\, Nonnenmacher, S.\,De Bi\`evre,
 Scarred eigenstates for quantum cat maps of minimal periods. {\em Comm. Math. Phys.} {\bf 239} (2003), no. 3, 449--492

\bibitem{GH}
 S.\,Gurevich and R.\,Hadani, "Proof of the Rudnick-Kurlberg Rate
 Conjecture", preprint


\bibitem{HB}
 J.H.\,Hannay and M.V.\,Berry, Quantization of linear maps on a
 torus --- Fresnel diffraction by a periodic grating, {\em Physica}
 D {\bf 1} (1980) 267-290.

\bibitem{KR}
 P.\,Kurlberg and Z.\,Rudnick, Hecke theory and equidistribution
 for the quantization of linear maps of the torus, {\em Duke Math. J.} {\bf 103} (2000) 47-77.

\bibitem{MO}
 J.Marklof , S. O'Keefe
 \newblock Weyl's law and quantum ergodicity for maps with divided phase
 space, preprint

\bibitem{MR}
 J.\,Marklof and Z.\,Rudnick, Quantum unique ergodicity for
 parabolic maps, {\em Geom. Funct. Anal.} {\bf 10} (2000)
 1554-1578.

\bibitem{S}
 W.\, M.\, Schmidt, Diophantine approximation Lecture notes in
 Mathematics, Vol. 785. Springer, Berlin, 1980

\bibitem{S1}
 W.\, M.\, Schmidt,  Approximation to algebraic numbers. Enseignement Math. (2) 17 (1971), 187--253.

 \bibitem{Shn}
 A.I. Snirelman, Ergodic properties of eigenfunctions, {\em Usp. Math. Nauk.} {\bf 29} (1974), 181-182.


\bibitem{Z1}
 S.\,Zelditch, Uniform distribution of eigenfunctions on compact
 hyperbolic surfaces, {\em Duke Math. J.} {\bf 55} (1987) 919-941.


\end{thebibliography}
\end{document}